\documentclass[reprint,longbibliography,
 amsmath,amssymb,
 prl,superscriptaddress]{revtex4-2}
\usepackage{graphicx}
\usepackage{dcolumn}
\usepackage{bm}
\usepackage[colorlinks=true, citecolor=blue, linkcolor=blue, urlcolor=blue]{hyperref}
\usepackage{xcolor}
\usepackage{amsmath, amssymb, amsfonts}
\usepackage{mathtools}
\usepackage{bbold}
\usepackage{subfigure}
\usepackage{mathrsfs}
\usepackage{sidecap}
\usepackage{verbatim}

\definecolor{purple1}{rgb}{128,0,128}

\usepackage{bigints}
\newcommand{\bea}{\begin{eqnarray}}
\newcommand{\ea}{\end{eqnarray}}

\definecolor{darkpastelgreen}{rgb}{0.01, 0.75, 0.24}

\def\sgn{\mbox{sgn}}

\def\d{\mathrm{d}}

\begin{document}

\title{Black hole analogues in dipolar condensates} 
\title{Black-hole radiation in dipolar condensates} 
\title{Impact of trans-Planckian excitations on black-hole radiation in dipolar condensates} 
\author{Caio C. \surname{Holanda Ribeiro}}
\affiliation{Seoul National University, Department of Physics and Astronomy, Center for Theoretical Physics, Seoul 08826, Korea} 
\affiliation{Institute of Physics, University of Brasilia, 70919-970 Brasilia, Federal District, Brazil
and International Centre of Physics, University of Brasilia, 70297-400 Brasilia, Federal District, Brazil} 
\author{Uwe R. Fischer}
\affiliation{Seoul National University, Department of Physics and Astronomy, Center for Theoretical Physics, Seoul 08826, Korea} 

\date\today

\begin{abstract}
We consider a quasi-one-dimensional dipolar condensate 
in an analogue black hole setup. It is shown that the existence of a roton minimum in the condensate dispersion relation leaves deep imprints onto the Hawking radiation spectrum.
In particular, the emitted radiation can be either more intense or suppressed, 
depending on the depth of  the roton minimum in the excitation spectrum. 
In addition, we find that spontaneous particle creation occurs even when the horizon is removed.
Our results establish that dipolar condensates offer a richer and more versatile environment for the simulation of  
particle production from the quantum vacuum in the presence of horizon-interfaces than their contact-interaction counterparts. 
\end{abstract}

\maketitle



{Black holes are a versatile laboratory to probe 
particle production from the quantum vacuum in the presence of horizon-interfaces separating distinct regions of spacetime  \cite{Hawking1975,PhysRevD.14.870,BroutReview}. 
A suitably clean and controllable arena to produce analogue black holes are Bose-Einstein condensates \cite{Lahav}, 
in which the first unambiguous detection of spontaneous quantum Hawking radiation has been achieved \cite{Steinhauer2019,Steinhauer2021}.
It has been argued in the years since its inception  
that Hawking radiation is a rather universal phenomenon, as the thermality of the Hawking spectrum is generally robust against trans-Planckian deformations of the spectrum breaking Lorentz invariance 
\cite{Unruh1995,Brout,Corley,Universality,ParentaniDispersive}, even though the emitted quanta at infinity, when traced back to the horizon, experience  an infinite blueshift \cite{Jacobson1991}. 
Black holes as well as their analogues are however 
conventionally set up in the field-theoretical context of contact interactions.
In the following, we demonstrate that admitting {\em nonlocal} field theories offer a much richer arena to harness 
the impact of trans-Planckian excitations on 
black-hole radiation. 
Specifically, we show that due to the increased complexity of the scattering problem at the horizon,  
novel features emerge which clearly distinguish dipolar black holes from their contact counterparts.
To this end, we take into account that interactions can be long-range and in particular anisotropic, for which dipole-dipole interactions
between atoms or molecules with magnetic or electric dipole moments are the archetype. Due to the roton minimum in their  elementary excitation spectrum \cite{PhysRevLett.90.250403,PhysRevA.73.031602,Giovanazzi2004,Chomaz2018}, 
black holes in dipolar condensates sensitively probe the robustness 
of Hawking radiation thermality to high frequency dispersion, thereby also thoroughly addressing one of 
the major original motivations of the whole analogue black hole program \cite{Unruh1981}.}

We thus provide below the first simulation of black hole (BH) analogues from dipolar Bose-Einstein condensates. Our system is assumed to be an elongated radially harmonic trapped flowing quasi-one-dimensional (quasi-1D) dipolar condensate, systems routinely realized in experiment \cite{Chomaz2022}. 
We assume that the system is stationary and sufficiently strongly elongated 
 such that the details of how the flow is sustained can be neglected to a first approximation 
 cf., e.g., Refs.~\cite{BLV2003,Parentani2009,Recati,Curtis,Pavloff2012,Boiron,Isoard,Balbinot2021,Balbinot2008}.

We recall that {three} ingredients are necessary to trigger spontaneous particle production {of quantum origin} 
in stationary condensates: The existence of negative energy excitations, 
a mechanism of 
mode conversion 
\cite{Jacobson1996,Parentani2009,Pavloff2012}, also cf.~Ref.~\cite{MattEssential}, 
{and that the fundamental commutation relations for the field in question are fulfilled \cite{UnruhSlow}}.
For contact BH analogues, negative energy superluminal excitations reach the event horizon from within the BH, which converts a part of these modes into outgoing spontaneous radiation, 
thus giving rise to the system's vacuum decay \cite{Parentani2009,Pavloff2012}. 
We show in what follows that spontaneous particle production also occurs in dipolar condensates when these 
{three} criteria are met. 
%
%

For quasi-1D dipolar condensates, {after integrating out the radial directions transverse to the long $x$ axis}, 
the ratio $\beta=\ell_\bot/\xi_{\rm u}$, 
where $\ell_\bot=1/\sqrt{\omega_\bot}$ is the transverse harmonic oscillator length
and $\xi_{\rm u}$ is the subsonic healing length outside the analogue BH 
(setting $\hbar=m=1$),  
measures how deep into the quasi-1D regime the system is \cite{Giovanazzi2004,Fischer2020}. There 
is a  critical $\beta_c\sim0.776$ for which the system becomes unstable against a proliferation of rotonic excitations
in the crossover to three spatial dimensions. 
In the limit $\beta\rightarrow 0$, {indicating a quasi-1D condensate with effective 
contact interactions, our model system coincides with the one 
studied in \cite{Curtis}, see for further details below.} 
At finite $\beta$, {for which the anisotropy of the dipolar interaction becomes manifest}, 
we find important differences between contact and dipolar BHs that can be summarized as follows: (i) $\beta>0$ increases the radiation power when the black hole exists; (ii) The rotonic/maxonic dispersion relation typical of dipolar gases leads to a strong nonthermality of the Hawking radiation spectrum in comparison to local black hole analogues; (iii) We find, due to the rotons, radiating systems even when a horizon is absent.
%
%
These findings demonstrate that dipolar condensates offer a much richer environment to simulate the Hawking phenomenon in comparison to contact condensates. 
         
Within the mean-field approximation, the evolution of the condensate order parameter $\phi$  is described by the nonlocal Gross-Pitaevskii equation (GPE) \cite{supplement}
\begin{equation}
i\partial_t\phi=\left[-\frac{1}{2}\partial_x^2+U+g_{\rm dd}|\phi|^2\right]\phi-3g_{\rm dd}\phi G*|\phi|^2,\label{nonlocalGP}
\end{equation}  
where $U=U(x)$ is the trap potential, and $g_{\rm dd}>0$ is the 
quasi-1D dipolar interaction strength.
{We assume that the dipoles have a common direction relative to the long $x$ axis, which fixes $g_{\rm dd}$ \cite{Fischer2020,Ribeiro2022}.} 
Note that we have separated off the contact part of the dipolar interaction (third term in square brackets)
 and we assume other contact interaction contributions coming from $s$-wave scattering to be negligible. This regime 
 can be achieved by using Feshbach resonances \cite{Chomaz2022}. 
For computational convenience, we {discretize}  
the quasi-1D dipolar kernel $G$ in Eq.~\eqref{nonlocalGP} \cite{supplement,Ribeiro2022}.
 Denoting the convolution by $G*f=\int dx'G(x-x')f(x')$, we have 

%
\begin{equation}
G(x)= \delta(x)-\frac{1}{2\ell_{\bot}}\frac{\sum_{j=1}^{\mathcal{N}}j^2\Delta q^3e^{-j^2\Delta q^2/2-j\Delta q |x|/\ell_{\bot}}}{\sum_{j=0}^{\mathcal{N}}j\Delta q^2e^{-j^2\Delta q^2/2}},\label{approxG}
\end{equation}
where the limit $\mathcal{N}\rightarrow\infty$, $\Delta q\rightarrow 0$ is understood, leading to the 
continuum expression $G(x)-\delta(x)=-(1/2\ell_{\bot})f(|x|/\ell_{\bot})$, $f(y)=-y+(1+y^2)\exp(y^2/2)\mbox{Erfc}(y/\sqrt{2})\sqrt{\pi/2}$, $\mbox{Erfc}$ being the complementary error function \cite{Fischer2020}. This form of writing the quasi-1D dipolar kernel enables the construction of essentially analytical 
solutions: 
We solve for finite $\mathcal{N}$ and $\Delta q$ and take the limits afterward. Also, by keeping $\mathcal{N}$ and $\Delta q$ finite we can produce approximate solutions to the problem to the desired accuracy. For instance, as shown in \cite{Ribeiro2022}, for $\mathcal{N}=10$ and $\Delta q=1/3.4$ the error 
is below $1\%$, 
maintained throughout our simulations. 

\begin{figure}[t!]
\includegraphics[scale=0.5]{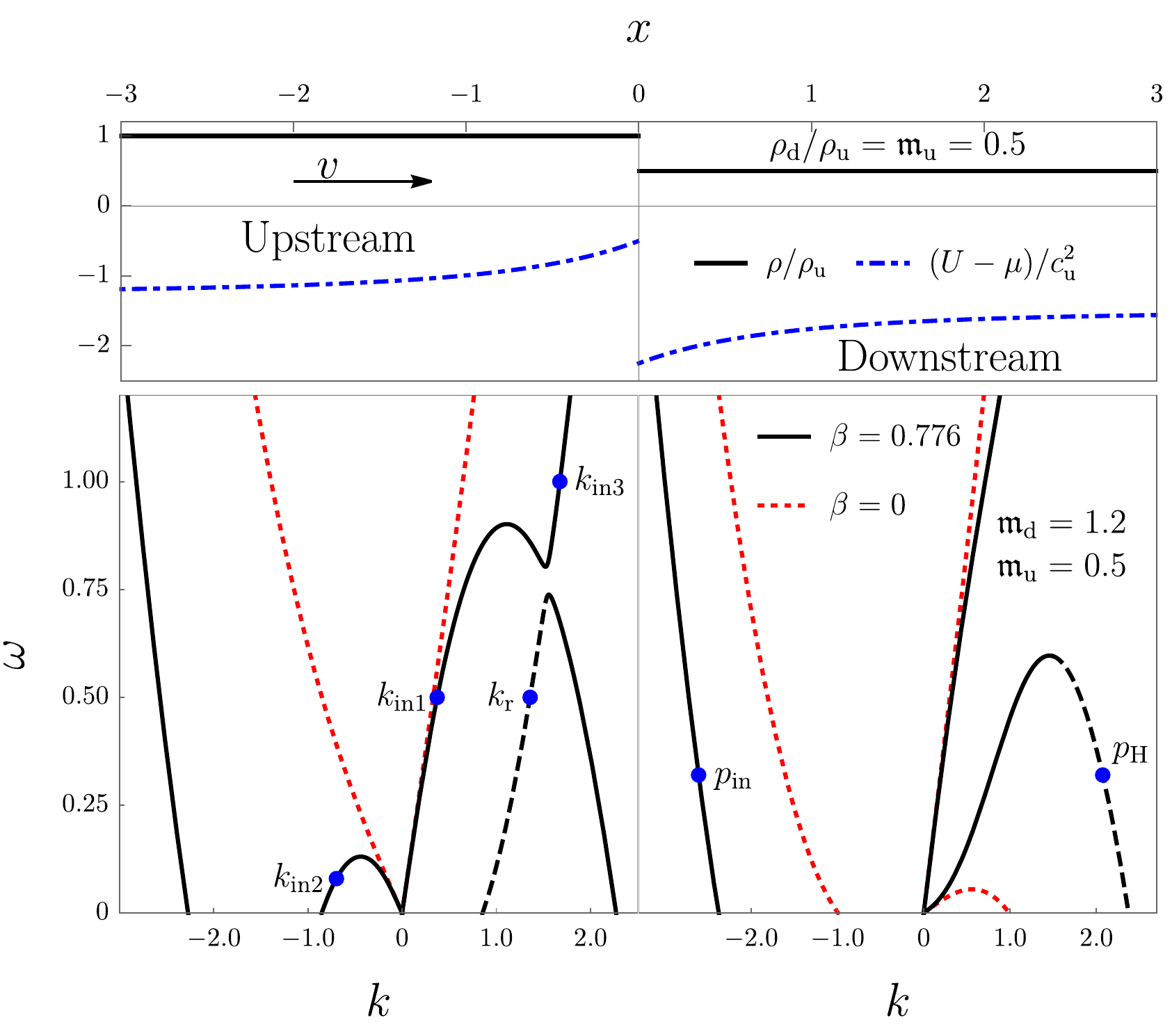}
\caption{Upper panel: Density jump model imposed by the external potential $U$. 
Lower panel: Bogoliubov dispersion relation for two BH analogues with distinct $\beta$ and fixed Mach numbers. Each line of constant $\omega$ intercepts the dispersion relation at the real wave vector solutions, {corresponding to 
plane waves propagating rightwards (leftwards) if the slope at the interception point is positive (negative)}. Left: Dispersion for the upstream region. Notice the strong deviation from a 
contact-interaction-dominated regime $\beta=0$ (red dotted curve), caused by the dipolar interactions. {Each plane wave propagating to the right gives rise to a distinct quasiparticle mode, indexed by $k_{\rm in1}$, $k_{\rm in2}$, $k_{\rm in3}$, and $k_{\rm r}$. Note that the ``rotonic'' branch $k_{\rm r}$ (dashed curve) has negative energy (negative norm), and 
that for contact-only BHs, only $k_{\rm in1}$ exists.} Right: Downstream dispersion relation. {Each leftwards propagating wave gives rise to a quasiparticle, indexed by $p_{\rm in}$ and the negative energy $p_{\rm H}$ (Hawking branch).} 
Dipolar interactions, in particular, increase the {cut-off} frequency of the local maximum on the right.
}
\label{figmain2}
\end{figure}

 {To convey the essential physics, we construct our stationary dipolar BH analogue as 
the solution $\phi=\sqrt{\rho}\exp(-i\mu t+ivx)$ ($v>0$) of Eq.~\eqref{nonlocalGP} for a piecewise constant density: $\rho=\rho_{\rm u}$ for $x<0$ and $\rho=\rho_{\rm d}<\rho_{\rm u}$ if $x>0$. 
A subscript on local quantities ``$\rm u$'' here and in what follows denotes the upstream region ($x<0$), and ``$\rm d$'' the downstream region $(x>0)$ (Fig.~\ref{figmain2} upper panel).}  
This model was recently studied for contact condensates in \cite{Curtis}. 
We assume zero temperature 
throughout. Equation \eqref{nonlocalGP} then implies continuity $\rho_{\rm u}v_{\rm u}=\rho_{\rm d}v_{\rm d}$ and $U=\mu+\partial_x^2\sqrt{\rho}/2\sqrt{\rho}-v^2/2-g_{\rm dd}\rho+3g_{\rm dd}G*\rho$
%
%
which fixes the external potential $U$ (cf.~Fig.~\ref{figmain2} upper panel). 

{We scale lengths in units of $\xi_{\rm u}$,~e.g.~$x=x[\xi_{\rm u}]$, 
so that the energy and (inverse time unit) becomes $1/\xi_{\rm u}^2$.}  A dipolar BH is then specified by {the set of parameters} $\{\mathfrak{m}_{\rm u},\mathfrak{m}_{\rm d}, \beta=\beta_{\rm u}=\ell_{\bot}\}$, where the local Mach number is $\mathfrak{m}=v/c$ and 
$c=\sqrt{g_{\rm dd}\rho}$ is the local sound speed. Then $\mathfrak{m}_{\rm u}<1<\mathfrak{m}_{\rm d}$ defines an analogue BH. 
Our goal 
is to determine how small fluctuations over this BH background lead to spontaneous radiation. 
The Bogoliubov expansion of the wave function reads $\hat{\Psi}=(\sqrt{\rho}+\hat{\psi})\exp(-i\mu t+ivx)$, where the 
bosonic operator 
$\hat{\psi}$ models the small quantum fluctuations. 

%
%

In order to study the Hawking radiation in our system, we expand the quantum field $\hat{\psi}$ in the unambiguous basis of quasiparticle modes whose vacuum state represents a zero flux of phonons sent towards the event horizon in the laboratory frame \cite{Parentani2009}. Hence for this vacuum choice any spontaneous radiation is linked to Hawking-like processes. Our representation for the interaction kernel \eqref{approxG} can be used to find such a basis 
as follows \cite{supplement}. We first define the Nambu spinor in particle-hole space, 
$\hat{\Phi}=
\left(\begin{smallmatrix}\hat{\psi}\\\hat{\psi}^\dagger\end{smallmatrix}\right) 
/\sqrt{\rho}$, 
and expand $
\hat{\Phi}=\sum_{n}(\hat{a}_{n}\Phi_n+\hat{a}^\dagger_{n}\sigma_1\Phi_n^*)$,
%
%
where $\{\Phi_n\}$ is a complete set of positive norm quasiparticle solutions with respect to the Bogoliubov scalar product: $\int dx\rho\Phi_n^\dagger\sigma_3\Phi_{n'}=\delta_{n,n'}$. The functions $\Phi_n$ are solutions of the 
Bogoliubov-de Gennes equation
\begin{align}
i\partial_t\sigma_3\Phi_n=&-\frac{1}{2\rho}\partial_x(\rho\partial_x\Phi_n)-iv\sigma_3\partial_x\Phi_n+\rho g_{\rm dd}\sigma_4\Phi_n\nonumber\\
&-3g_{\rm dd}\sigma_4G*\Phi_n.\label{bogo}
\end{align}
We find that $\sigma_1 {\Phi}^*_n$ is a negative norm solution, 
and $\sigma_i$, $i=1,2,3$ are Pauli matrices, with $\sigma_4=1+\sigma_1$. 

Solutions to Eq.~\eqref{bogo} can be found with the time dependence $\exp(-i\omega t)$ 
for $\omega>0$. Furthermore, because Eq.~\eqref{approxG} is suppressed for $|x|\gg 1,\ell_{\bot}$,
all solutions of Eq.~\eqref{bogo} far from the horizon are written in terms of the local homogeneous condensate perturbations, i.e., we have superposition of plane waves in the form $\Phi(t,x)=\exp(-i\omega t +ikx)\Phi_k$ for constant $\Phi_k$. This gives rise to the dispersion relation 
\begin{equation}
\omega=\mathfrak{m}_{\rm u}\frac{\rho_{\rm u}}{\rho}k\pm k\sqrt{(\rho/\rho_{\rm u})[1-3\tilde{G}(\beta k)]+k^2/4},\label{disp}
\end{equation}
where $\tilde{G}(\beta k)=\tilde{G}(\ell_\bot k)=\int dx G(x)\exp(-i\ell_\bot kx/\ell_\bot)$.

Equation \eqref{disp} 
enables us to identify a quasiparticle basis whose vacuum state is by definition characterized by no quasiparticles  propagating towards the event horizon. This state is, therefore, suitable for studying the spontaneous Hawking radiation in BH analogues. 
%
A positive (negative) sign of the dimensionless group velocity $V(k)=\d\omega/dk$ (sign of the slopes in Fig.~\ref{figmain2} lower panel) determines whether the plane wave is propagating to the right (left), and each plane wave propagating towards the event horizon gives rise to a distinct quasiparticle mode found by solving the scattering problem within Bogoliubov theory \cite{Parentani2009,Pavloff2012,Parentani_2016-2,Ribeiro2022}. 

Let us recollect the salient features of the dispersion relation in the contact BH regime (dotted curves of Fig.~\ref{figmain2} bottom). When $\beta=0$ there is only one plane wave going towards the horizon from the upstream region (Fig.~\ref{figmain2} bottom left) for each $\omega$, whereas in the downstream region (Fig.~\ref{figmain2} bottom right) for frequencies below the local maximum (Hawking cut-off frequency) there are always two plane waves propagating towards the horizon, with the negative energy quasiparticles indexed by $p_{\rm H}$ from the ``Hawking branch'' (dashed curve in Fig.~\ref{figmain2} bottom right) leading to the spontaneous radiation process \cite{Pavloff2012}. When $\beta>0$, the observed effect inside the black hole is the increase of the Hawking frequency, suggesting a stronger radiation power \cite{Pavloff2012}. Furthermore, novel phenomena are expected to occur outside the black hole. As $\beta$ continuously increases from zero, initially no qualitative 
distinction from the contact case occurs {(dotted curves in Fig.~\ref{figmain2} bottom)}.  
However, when the roton minimum emerges (continuous curves in Fig.~\ref{figmain2} bottom), the new plane wave branch {$k_{\rm in2}$} approaches the horizon {from the upstream region}. This was recently shown to have a great impact on the scattering process 
at interfaces \cite{Ribeiro2022}, and thus also affects the Hawking radiation {as we show in the below}. Finally, if $\beta$ is increased even further (Fig.~\ref{figmain2} bottom left, dashed curve), beyond {$\beta=\beta_{\rm r}$ 
($\simeq 0.628$}  for $\mathfrak{m}_{\rm u}=1/2$),   
a novel negative energy quasiparticle branch {indexed by $k_{\rm r}$} emerges from the upstream region, which we call the ``rotonic branch'' for simplicity. 
The occurrence of the latter is indicative of a strong departure from a contact-dominated BH. Indeed, whereas the 
{maximal number of quasiparticles at given frequency} is $3$ in the contact case, for the dipolar BH we find up to $6$ quasiparticles, depending on $\beta$ (cf. Fig.~\ref{figmain2} bottom, showing that different mechanisms of mode conversion take place for this dipolar BH. Therefore, the scattering problem in our dipolar case is significantly more intricate than for the contact BH analogues. 

%
%

{We show in the supplement \cite{supplement} that the normalized quasiparticles for $\omega>0$ can be written as
\begin{align}
\Phi^{(\alpha)}_\omega=e^{-i\omega t}\left\{
\begin{array}{c}
\sum_{p}S_{p}^{(\alpha)}e^{ipx}\Phi_{p},\ x>0,\\
\sum_{k}S_{k}^{(\alpha)}e^{ikx}\Phi_{k},\ x<0,
\end{array}\right.\label{gensol}
\end{align}
where we denote the downstream wave vectors by $p$. The index $\alpha$ assumes values in the two sets $\Gamma^{(+)}=\{k_{\rm in1},k_{\rm in2},k_{\rm in3},p_{\rm in}\}$ and $\Gamma^{(-)}=\{k_{\rm r},p_{\rm H}\}$, and for each $\alpha$, the sums in Eq.~\eqref{gensol} include the incoming channel with $S^{(\alpha)}_{\alpha}=1$, all outgoing propagating channels, and all evanescent waves. Both $S_{k}^{(\alpha)}$ and $S_{p}^{(\alpha)}$ are fixed by Eq.~\eqref{bogo}, and the sign of the norm 
\begin{eqnarray}
\Phi_{k}^\dagger\sigma_3\Phi_{k}=\frac1{2\pi \rho |V(k)|}\mbox{sgn}(\omega-\mathfrak{m}_{\rm u}k\rho_{\rm u}/\rho)
\end{eqnarray}  
determines whether an incoming mode propagating towards the horizon has positive 
or negative energy  for real $k$ \cite{supplement}. Furthermore, the quasiparticles in $\Gamma^{(+)}$ have positive norm, whereas the ones in $\Gamma^{(-)}$ have negative norm. Accordingly, the full field operator expansion reads 
\begin{align}
\hat{\Phi}=&\int_0^\infty\d\omega\Bigg[ \sum_{\alpha\in\Gamma^{(+)}}(\hat{a}^{(\alpha)}_{\omega}\Phi^{(\alpha)}_\omega+\hat{a}^{(\alpha)\dagger}_{\omega}\sigma_1\Phi^{(\alpha)*}_\omega)\nonumber\\
&+\sum_{\alpha\in\Gamma^{(-)}}(\hat{a}^{(\alpha)\dagger}_{\omega}\Phi^{(\alpha)}_\omega+\hat{a}^{(\alpha)}_{\omega}\sigma_1\Phi^{(\alpha)*}_\omega)\Bigg]. \label{PhiExpansion} 
\end{align} 
We adopt  the convention that for each index $\alpha$, both $\hat{a}^{(\alpha)}_{\omega}$ and $\Phi^{(\alpha)}_\omega$ are zero if the pair $(\omega,\alpha)$ does not index a solution of the BdG equation. 
This greatly simplifies the notation, 
as it allows us to write the field expansion as a single $\omega$ 
integral from $0$ to $\infty$.  Furthermore,  $[\hat{a}^{(\alpha)}_{\omega},\hat{a}^{(\alpha')\dagger}_{\omega'}]=\delta_{\alpha,\alpha'}\delta(\omega-\omega')$, and the vacuum state $|0\rangle$ is defined by 
$\hat{a}^{(\alpha)}_{\omega}|0\rangle=0$.}

\begin{figure}[t]
\includegraphics[scale=0.56]{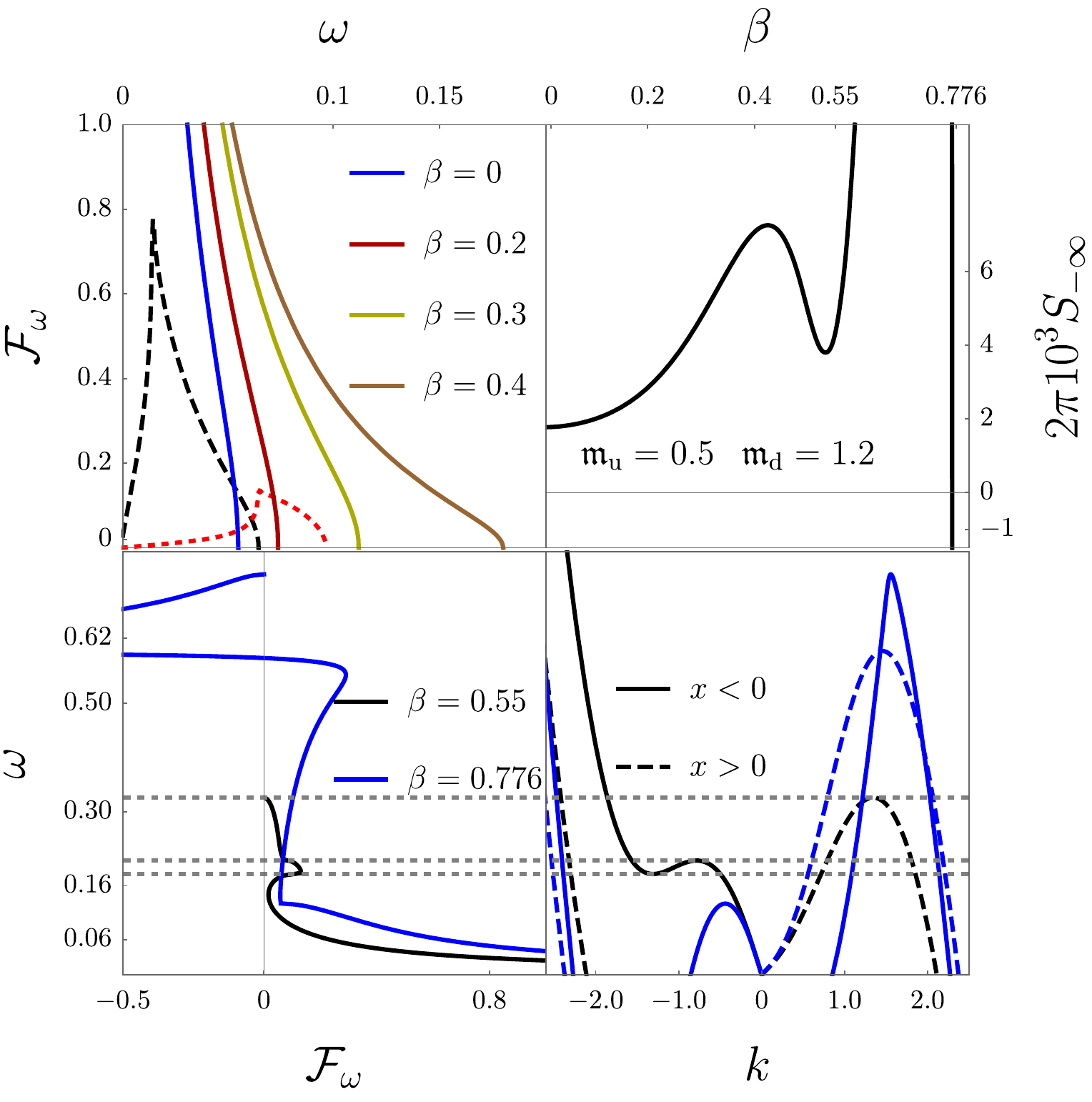}
\caption{Top left: Spectral distribution of radiation $\mathcal{F}_{\omega}$ in Eq.~\eqref{rpower},  
for a dipolar BH  analogue ($\mathfrak{m}_{\rm d}=1.2, \mathfrak{m}_{\rm u}=0.5$) and for several values of $\beta$. The blue curve for $\beta=0$ corresponds to a contact BH. 
Dashed and dotted curves are the radiation spectrum for two non-BH 
configurations with $\mathfrak{m}_{\rm d}=0.9, \beta=0.5$ and $\mathfrak{m}_{\rm d}=0.6, \beta=0.65$, respectively. 
Top right: Radiation power as function of $\beta$. 
As $\beta$ increases from zero, $S_{-\infty}$ increases, achieves a local minimum at about $\beta\sim0.55$ and  eventually becomes negative when the system approaches the deep roton minimum  (close to $\beta_c\sim0.776$), 
for which the radiation direction is reversed, indicating the proximity to instability 
of the quasi-1D dipolar gas, and a strong departure from thermal radiation is observed (bottom left). 
The local minimum in $S_{-\infty}$ corresponds to the appearance  of upstream quasiparticle branches with small group velocities (bottom right). 
In the bottom part, we also display the correlation between spectral distribution for $\beta=0.55$ and $0.776$ (left) 
{and the dispersion relation branch with negative sign in Eq.~\eqref{disp} (right), from which the rotonic $k_{\rm r}$ and 
the Hawking $p_{\rm H}$ modes originate, cf.~Fig.~\ref{figmain2}, lower panel.}
} 
\label{figmain3}
\end{figure}

The operator $\hat{\psi}$ is the 
upper component of $\sqrt{\rho}\hat{\Phi}$, and once the quantum field expansion is obtained,  
we find that the (normal ordered) system Hamiltonian assumes the diagonal form \cite{supplement}
\begin{align}
\hat{H}=\int_0^\infty\d\omega \omega \left[\sum_{\alpha\in\Gamma^{(+)}}\hat{a}^{(\alpha)\dagger}_{\omega}\hat{a}^{(\alpha)}_{\omega}-\sum_{\alpha\in\Gamma^{(-)}}\hat{a}^{(\alpha)\dagger}_{\omega}\hat{a}^{(\alpha)}_{\omega}\right]
\end{align} 
similar to the contact BH \cite{Parentani2009}. The Hamilton operator above demonstrates that exciting a quasiparticle mode with index in $\Gamma^{(-)}$ diminishes the system energy. We can calculate the corresponding radiation spectrum deep into the upstream region ($x\rightarrow-\infty$) 
via the following evolution equation for the Hamiltonian density $\mathcal{\hat{H}}$  
\cite{supplement},   
\begin{align}
&\partial_t\mathcal{\hat{H}}=-\partial_x\hat{S} 
+\frac{3\sqrt{\rho}g_{\rm dd}}{2}\left\{(\partial_t\hat{\psi})G*\sqrt{\rho}\hat{\psi}-\hat{\psi}G*\sqrt{\rho}\partial_t\hat{\psi}
\right.\nonumber\\
&+\left.[G*(\sqrt{\rho}\hat{\psi}^\dagger)]\partial_t\hat{\psi}-\hat{\psi}^\dagger G*\sqrt{\rho}\partial_t\hat{\psi}+\mbox{H.c.}\right\}.
\end{align}
where $\hat{S}=-(\partial_t\hat{\psi}^\dagger)(\partial_x-iv)\hat{\psi}/2+\mbox{H.c.}$ and $\mathcal{\hat{H}}$ is the Hamiltonian density. Notice that only when $\beta=0$ and thus $G=0$ (contact-only case) energy is locally conserved in the system, whereas for any finite $\beta$ there is no local energy conservation. Nevertheless, we find generally that energy is globally conserved \cite{supplement}: $\partial_tH=-S_{\infty}+S_{-\infty}=0$, 
%
%
$H=\int \d x\langle\mathcal{\hat{H}}\rangle$ is the system energy, and the outgoing flux becomes the following $\omega$-integral
%
\begin{align}
S_{-\infty}&= \frac{1}{2\pi} 
\int_{0}^{\infty} \d\omega \omega \mathcal{F}_{\omega},\nonumber
\\
%
%
\mathcal{F}_{\omega}&=\sum_{\alpha\in\Gamma^{(-)}}\sum_{k\ {\rm prop}}|S^{(\alpha)}_{k}|^2\sgn\left[V(k)(\omega-\mathfrak{m}_{\rm u}k)\right]. \label{rpower}
\end{align}
%
The sum in $k$ is performed over all real upstream propagating 
wave vectors only. The quantity $S_{-\infty}$ is then identified as the power radiated at $x\rightarrow-\infty$, and $\mathcal{F}_{\omega}$ 
is the corresponding power spectrum, containing the negative
energy modes from Eq.~\eqref{PhiExpansion}.  
We depict $\mathcal{F}_{\omega}$ in Fig.~\ref{figmain3} top left panel for several values of $\beta$ and $\mathfrak{m}_{\rm d}=1.2, \mathfrak{m}_{\rm u}=0.5$. 
For the contact BH analogue $\beta=0$ the spectrum is approximately thermal in the low-frequency regime \cite{Pavloff2012}. We note that as $\beta$ increases from zero ($\beta=0.2, 0.3, 0.4$ in Fig.~\ref{figmain3} top left), the effect of the dipolar interactions is to increase the spectrum cut-off frequency and the radiated power (Fig.~\ref{figmain3} top right). Dipolar black holes therefore represent a more promising scenario to probe Hawking radiation 
from the point of view of the radiated energy than local condensates.
For larger values of $\beta$, the emergence of new quasiparticle branches with excitations of low group velocity then 
leads to a decrease in the radiated power, see Fig.~\ref{figmain3} top right and bottom panels. As $\beta$ approaches the deep roton regime 
and the rotonic quasiparticle branch {($k_{\rm r}$)} appears beyond $\beta_{\rm r}$, a marked distinction of the dipolar BH when compared to contact BHs is observed, as two (negative energy) mode conversion mechanisms {originating in  the quasiparticles $k_{\rm r}$ and $p_{\rm H}$} compete in the radiation process. Figure \ref{figmain3} top right and bottom panels show that the rotonic branch can, in principle, even suppress the radiation power or revert its direction, which represents an instance of
significant impact of trans-Planckian physics.

In Fig.~\ref{figmain3} top left panel we {also} depict the radiation spectrum for two non-BH configurations, the dashed and dotted curves corresponding to a flow with $\mathfrak{m}_{\rm d}=0.9, \beta=0.5$ and $\mathfrak{m}_{\rm d}=0.6, \beta=0.65$, respectively. We recall that spontaneous particle creation can occur if both negative energy excitations exist and a mechanism that converts such negative energy excitations into outgoing radiation 
 is present. In the flowing dipolar condensate, the negative energy rotonic {and Hawking-type} quasiparticles might be present independently of the value of $\mathfrak{m}_{\rm d}$, and can be scattered
 at the interface at $x=0$ owning to outgoing radiation even without a horizon ($\mathfrak{m}_{\rm d}<1$), leading to 
 the (strongly nonthermal) radiation spectrum presented in Fig.~\ref{figmain3} top left panel. 

In conclusion, we considered here for the first time the possibility of simulating BHs in 
dipolar Bose-Einstein condensates. Our analysis shows that the presence of dipolar interactions leads to a marked departure from contact 
condensates, including a novel mechanism of mode conversion of the rotonic branch of quasiparticles at the interface, 
leading to radiating scenarios even when there is no horizon.  
{The presence of a roton minimum leads to strong nonthermality of the spectrum as
in the analogue Unruh effect \cite{Zehua},  and to enhanced radiation power, making dipolar condensates 
especially promising candidates to probe Hawking radiation.  
Generally, we expect in dipolar gases significant alterations 
from the predictions of a Lorentz-invariant theory, also 
and in particular in cosmological scenarios which employ a similar dispersion relation see, e.g., 
\cite{Starobinsky,Lemoine,PhysRevD.89.043507,PhysRevLett118.130404}.} 
{Finally, our analysis focuses on the existence of quasiparticle radiation mechanisms which characterize a dipolar BH analogue model. 
Future studies will investigate the effects of ``zero modes'' on the very dipolar condensate existence 
\cite{Lewenstein,Wang,Isoard}, and the radiation 
and pair-entanglement verification  
procedure via density-density correlations \cite{Balbinot2008,Finazzi_2014,Steinhauer2015,Parentani_2017},} 


%
%

This work has been supported by the National Research Foundation of Korea under 
Grants No.~2017R1A2A2A05001422 and No.~2020R1A2C2008103.

\bibliography{DBH10}

%
%

\vspace*{70em} 
\newpage
\begin{widetext}
\setcounter{equation}{0}
\setcounter{figure}{0}
\setcounter{table}{0}
\setcounter{page}{1}
\renewcommand{\theequation}{S\arabic{equation}}
\renewcommand{\thefigure}{S\arabic{figure}}

\section{Supplemental Material}

\subsection{Bogoliubov-de Gennes equation}

We consider a near-quasi-1D dipolar condensate modeled by the order parameter $\phi$ solution of the nonlocal GPE%
\begin{equation}
i\partial_t\phi=\left(-\frac{\partial_x^2}{2}+U+g_{\rm dd}\rho\right)\phi-3g_{\rm dd}\phi G*\rho,\label{sm1}
\end{equation}
where $\rho=|\phi|^2$ is the condensate density. Here, the symbol $G*\rho$ indicates the convolution $G*\rho(x)=\int_{-\infty}^{\infty} dx'G(x-x')\rho(x')$, where the interaction kernel $G$ is defined in terms of its Fourier transform by $G(x)=(1/2\pi)\int dk \tilde{G}(\ell_{\bot}k)\exp(ikx)$, and $\tilde{G}(\ell_{\bot}k)$ is the limit $\mathcal{N} \rightarrow\infty$, $\Delta q\rightarrow 0$ of
\begin{equation}
\tilde{G}(\eta)= \frac{\eta^2}{\sum_{j=0}^{\mathcal{N}}j\Delta q^2e^{-j^2\Delta q^2/2}}\sum_{j=0}^{\mathcal{N}}\frac{j\Delta q^2e^{-j^2\Delta q^2/2}}{j^2\Delta q^2+\eta^2}.\label{sm2}
\end{equation}
The advantage of writing the interaction kernel as in the equation above comes from the fact that for $\mathcal{N}\sim10$ and $\Delta q\sim1/3.4$ one already has an excellent approximation to the exact kernel. Also, $\ell_{\bot}$ is the characteristic radial size o the condensate. Equation \eqref{sm1} can also be expressed using the Madelung representation $\phi=\sqrt{\rho}\exp(i\theta)$ in terms of the system density and phase $\theta$ as
\begin{align}
\partial_t\rho&=-\partial_x(\rho v),\label{sm3}\\
-\partial_t\theta&=-\frac{\partial_x^2\sqrt{\rho}}{2\sqrt{\rho}}+\frac{v^2}{2}+U+g_{\rm dd}\rho-3g_{\rm dd}G*\rho,\label{sm4}
\end{align}
where $v=\partial_x \theta$. Equation \eqref{sm3} is just the continuity equation, and Eq.~\eqref{sm4} is a nonlocal, dipolar generalization of the Euler equation. Furthermore, the GPE follows from the Lagrangian
\begin{equation}
L=\int dx\left[\frac{i}{2}\phi^*\partial_t\phi-\frac{i}{2}(\partial_t\phi^*)\phi-\frac{1}{2}|\partial_x\phi|^2-\left(U+\frac{g_{\rm dd}}{2}|\phi|^2\right)|\phi|^2+\frac{3g_{\rm dd}}{2}|\phi|^2G*|\phi|^2\right].
\end{equation} 

Given a solution $\phi$ of Eq.~\eqref{sm1}, we want to study small perturbations of the form $\phi\rightarrow\phi+\delta\phi$, that correspond to $L\rightarrow L+\delta L$, where
\begin{align}
\delta L=\int dx\bigg[&\frac{i}{2}\delta\phi^*\partial_t\delta\phi-\frac{i}{2}(\partial_t\delta\phi^*)\delta\phi-\frac{1}{2}|\partial_x\delta\phi|^2-\left(U+2g_{\rm dd}\rho-3g_{\rm dd}G*\rho\right)|\delta\phi|^2-\frac{g_{\rm dd}}{2}(\phi^2\delta\phi^{*2}+\phi^{*2}\delta\phi^{2})\nonumber\\
&+\frac{3g_{\rm dd}}{2}(\phi^*\delta\phi+\phi\delta\phi^*)G*(\phi^*\delta\phi+\phi\delta\phi^*)\bigg],
\end{align}  
in view of the GPE. Thus, by starting from the solution $\phi=\sqrt{\rho}\exp(-i\mu t+ivx)$ to the GPE, and defining our field variable $\psi$ by $\delta\phi=\exp(-i\mu t+ivx)\psi$, we obtain the Lagrangian $L_\psi$ for the field $\psi$
\begin{align}
L_{\psi}=\int dx\bigg\{&\frac{i}{2}\psi^*\partial_t\psi-\frac{i}{2}(\partial_t\psi^*)\psi-\frac{1}{2}|\partial_x\psi|^2-\left(g_{\rm dd}\rho+\frac{\partial_x^2\sqrt{\rho}}{2\sqrt{\rho}}\right)|\psi|^2-\frac{g_{\rm dd}}{2}\rho(\psi^{*2}+\psi^{2})\nonumber\\
&+\frac{iv}{2}\left[\psi^*\partial_x\psi-(\partial_x\psi^*)\psi\right]+\frac{3g_{\rm dd}}{2}\sqrt{\rho}(\psi+\psi^*)G*\sqrt{\rho}(\psi+\psi^*)\bigg\},\label{lagrangian}
\end{align}  
and the Euler-Lagrangian equation
\begin{equation}
i\partial_t\psi=-\frac{\partial_x^2}{2}\psi-iv\partial_x\psi+\left[\frac{\partial_x^2\sqrt{\rho}}{2\sqrt{\rho}}-\frac{i}{2}(\partial_xv)\right]\psi+\rho g_{\rm dd}(\psi+\psi^*)-3g_{\rm dd}\sqrt{\rho}G*\sqrt{\rho}(\psi+\psi^*),\label{BdG}
\end{equation}
known as the Bogoliubov-de Gennes equation. In order to find the solutions of Eq.~\eqref{BdG}, it is convenient to work with the Nambu spinor defined by
\begin{equation}
\Psi=\frac{1}{\sqrt{\rho}}
\left(\begin{array}{c}
\psi\\
\psi^*
\end{array}\right),
\end{equation}
which is seen to satisfy the reflection property $\sigma_1\Psi^*=\Psi$, and the BdG equation in the form
\begin{equation}
i\partial_t\sigma_3\Psi=-\frac{1}{2\rho}\partial_x\left(\rho\partial_x\Psi\right)-iv\sigma_3\partial_x\Psi+\rho g_{\rm dd}\sigma_4 \Psi-3g_{\rm dd}\sigma_4G*\rho\Psi.\label{BdGtosolve}
\end{equation}
Here, $\sigma_i$, $i=1,2,3$ are the Pauli matrices, and $\sigma_4=1+\sigma_1$. Solutions of the equation above are such that $\Psi$, $\rho\partial_x \Psi$ are everywhere continuous functions, properties used in our model as boundary conditions at the event horizon. Equation \eqref{BdGtosolve} implies that the quantity
\begin{equation}
\langle\Psi,\Psi'\rangle=\int dx\rho\Psi^\dagger\sigma_3\Psi',
\end{equation}
is conserved in time, which we use as a scalar product on the space of solutions to \eqref{BdGtosolve}. 

\subsection{Field modes {and the scattering problem}}

Solutions to Eq.~\eqref{BdGtosolve} can be found in the form $\Psi(t,x)=\exp(-i\omega t)\Psi_{\omega}(x)$, $\omega>0$, where
\begin{equation}
\omega\sigma_3\Psi_{\omega}=-\frac{1}{2\rho}\partial_x\left(\rho\partial_x\Psi_{\omega}\right)-i\frac{\rho_{\rm u}}{\rho}\mathfrak{m}_{\rm u}\sigma_3\partial_x\Psi_{\omega}+\frac{\rho}{\rho_{\rm u}} \sigma_4 \Psi_{\omega}-\frac{3}{\rho_{\rm u}}\sigma_4G*\rho\Psi_{\omega}.\label{BdGtosolve2}
\end{equation}
For contact condensates, the solutions to Eq.~\eqref{BdGtosolve2} for $x\neq0$ are superpositions of plane waves. When (nonlocal) dipolar interactions are present, however, that is not the case, as shown in \cite{Ribeiro2022}. Still, our representation for the dipolar kernel \eqref{approxG} is such that the solutions to the problem resemble superpositions of plane waves, and can be found as follows. For $|x|\gg 1,\ell_{\bot}$ {(we remind the reader that we scale lengths in units of $\xi_{\rm u}$)}, because $G(x)\rightarrow0$ when $x\rightarrow\infty$, any solution of Eq.~\eqref{BdGtosolve2} becomes a combination of $\exp(ikx)\Phi_k$, for constant $\Phi_k$, such that
\begin{equation}
\omega\sigma_3\Phi_k=\frac{k^2}{2}\Phi_k+k\frac{\rho_{\rm u}}{\rho}\mathfrak{m}_{\rm u}\sigma_3\Phi_k+\frac{\rho}{\rho_{\rm u}} \sigma_4 [1-3\tilde{G}(\beta k)]\Phi_k.
\end{equation}
Hence non-trivial solutions for $\Phi_k$ exist only if the dispersion relation holds
\begin{equation}
\left(\omega-k\frac{\rho_{\rm u}}{\rho}\mathfrak{m}_{\rm u}\right)^2=k^2\left\{\frac{\rho}{\rho_{\rm u}}[1-3\tilde{G}(\beta k)]+\frac{k^2}{4}\right\},
\end{equation}
fixing possible $\Phi_k$ solutions for each $\omega$. The group velocity $V(k)=d\omega/d k$ thus indicate if the plane wave is propagating to the right or leftwards. For the case of dipolar condensate, the dispersion relation admits in general more wave vector solutions for each $\omega$ in comparison to local BH analogues \cite{Ribeiro2022}. We showed in \cite{Ribeiro2022} that each plane wave propagating towards (the event horizon at) $x=0$ combined with transmitted, reflected and evanescent waves gives rise to a quasiparticle mode 
%
%
%
\begin{align}
\Psi^{(\alpha)}_{\omega}(x)=\left\{
\begin{array}{c}
\sum_{p}S_{p}^{(\alpha)}e^{ipx}\Phi_{p},\ x>0,\\
\sum_{k}S_{k}^{(\alpha)}e^{ikx}\Phi_{k},\ x<0,
\end{array}\right.\label{supp1}
\end{align}
where we denoted by $p$ the wave vector solutions for $x>0$, and the $\alpha$ index distinct 
{ingoing (propagating towards the horizon)} quasiparticles for a given $\omega$, {i.e., $\alpha\in\{k_{\rm in1},k_{\rm in2},k_{\rm in3}, k_{\rm r},p_{\rm in},p_{\rm H}\}$}. Also, 
\begin{align}
\Phi_{k}=\left|\frac{k^2}{4\pi \rho V(k)(\omega-\mathfrak{m}_{\rm u}k\rho_{\rm u}/\rho)(\omega-\mathfrak{m}_{\rm u}k\rho_{\rm u}/\rho-k^2/2)^2}\right|^{1/2}\left(\begin{array}{c}
(1-3\tilde{G})\rho/\rho_{\rm u}\\
\omega-\mathfrak{m}_{\rm u}k\rho_{\rm u}/\rho-k^2/2-(1-3\tilde{G})\rho/\rho_{\rm u}
\end{array}\right).\label{suppnorm}
\end{align}
Furthermore, {we set $S^{(k_{\rm in1})}_{k_{\rm in1}}=S^{(k_{\rm in2})}_{k_{\rm in2}}=S^{(k_{\rm in3})}_{k_{\rm in3}}=S^{(k_{\rm r})}_{k_{\rm r}}=S^{(p_{\rm in})}_{p_{\rm in}}=S^{(p_{\rm H})}_{p_{\rm H}}=1$ to have ``unit'' signals approaching the horizon, and the sums in Eq.~\eqref{supp1} include, in addition to the incoming channel $\alpha$, all the outgoing propagating channels, and evanescent waves solutions of the dispersion relation. By counting the number of $S^{(\alpha)}_{k}$, $S^{(\alpha)}_{p}$, we find $4+2\mathcal{N}$ unknowns for each $\mathcal{N}$ \cite{Ribeiro2022}.}
 {The various coefficients $S^{(\alpha)}_k$, $S^{(\alpha)}_p$ are fixed by the $2\mathcal{N}$ conditions
%
%
\begin{equation}
\sum_{k}\frac{S^{(\alpha)}_k}{k-ij\Delta q/\beta}\sigma_4\Phi_{k}=\sum_{p}\frac{S^{(\alpha)}_p}{p-ij\Delta q/\beta}\sigma_4\Phi_{p},
\end{equation}
for $-\mathcal{N}\leq j\leq\mathcal{N}$, $j\neq0$, plus the $4$ boundary conditions: $\Psi^{(\alpha)}_{\omega}$, $\rho\partial_x \Psi^{(\alpha)}_{\omega}$ continuous at $x=0$ \cite{Ribeiro2022}, such that the solutions to the BdG equation are found.  
}

{For each quasiparticle $\alpha$, to find all the (scattering) coefficients $S^{(\alpha)}_k$, $S^{(\alpha)}_p$ for real $k$, $p$ is a problem known as the Scattering Problem, which amounts to determine how the interface at $x=0$ scatters plane waves sent towards it. An important concept in such analysis is that of unitarity. The latter is stated in terms of constraints satisfied by the various scattering coefficients as follows. Equation \eqref{BdGtosolve2} implies that
\begin{align}
\rho(\omega'-\omega)\Psi^{(\alpha)\dagger}_{\omega}\sigma_3\Psi^{(\alpha')}_{\omega'}=&\frac{\partial_x}{2}
\rho\left\{\Psi^{(\alpha)\dagger}_{\omega}\partial_x\Psi^{(\alpha')}_{\omega'}-[\partial_x\Psi^{(\alpha)\dagger}_{\omega}]\Psi^{(\alpha')}_{\omega'}+2i\mathfrak{m}_{\rm u}\frac{\rho_{\rm u}}{\rho}\Psi^{(\alpha)\dagger}_{\omega}\sigma_3\Psi^{(\alpha')}_{\omega'}\right\}\nonumber\\
&+\frac{3\rho}{\rho_{\rm u}}\left[\Psi^{(\alpha)\dagger}_{\omega}\sigma_4G*\rho\Psi^{(\alpha')}_{\omega'}-(G*\rho\Psi^{(\alpha)\dagger}_{\omega})\sigma_4\Psi^{(\alpha')}_{\omega'}\right]:=I^{\alpha,\alpha'}_{\omega,\omega'},\label{cons1}
\end{align}
and the orthogonality is translated as $\int dx I^{\alpha,\alpha'}_{\omega,\omega'}=0$ for all $\omega$, $\omega'$. By performing this integral and making the substitution $\omega'\rightarrow\omega$, we obtain the aforementioned constraint
\begin{equation}
\sum_{k\ {\rm prop}}S^{(\alpha')}_{k}S^{(\alpha)*}_{k}\sgn\left[V(k)(\omega-\mathfrak{m}_{\rm u}k)\right]-\sum_{p\ {\rm prop}}S^{(\alpha')}_{p}S^{(\alpha)*}_{p}\sgn\left[V(p)(\omega-\mathfrak{m}_{\rm u}p\rho_{\rm u}/\rho_{\rm d})\right]=0,\label{smatrix}
\end{equation}
for all $\alpha$, $\alpha'$.
}

{The} normalization of Eq.~\eqref{suppnorm} and Eq.~\eqref{smatrix} ensure that $\langle\Psi_\omega^{(\alpha)},\Psi_{\omega'}^{(\alpha')}\rangle=\pm\delta_{\alpha,\alpha'}\delta(\omega-\omega')$, where $+$ and $-$ signs
stand for positive and negative norm
modes, respectively.  
We let $\Gamma^{(+)}$ (resp.~$\Gamma^{(-)}$) be the index set for positive (resp.~negative) norm quasiparticle modes, and the field modes become $\Phi^{(\alpha)}_\omega(t,x)=\exp(-i\omega t)\Psi^{(\alpha)}_{\omega}(x)$. {After a lengthy manipulation, we find that $\Gamma^{(+)}=\{k_{\rm in1},k_{\rm in2},k_{\rm in3},p_{\rm in}\}$, and $\Gamma^{(-)}=\{k_{\rm r},p_{\rm H}\}$.}  Furthermore, if {the indices $\alpha,\omega$ are such that $\Phi^{(\alpha)}_\omega(t,x)$ is not a solution to the BdG equation, we define $\Phi^{(\alpha)}_\omega(t,x)=0$. With this notation, the quantum field expansion then reads} 
%
\begin{align}
\hat{\Phi}=&\int_0^\infty\d\omega\Bigg[ \sum_{\alpha\in\Gamma^{(+)}}(\hat{a}^{(\alpha)}_{\omega}\Phi^{(\alpha)}_\omega+\hat{a}^{(\alpha)\dagger}_{\omega}\sigma_1\Phi^{(\alpha)*}_\omega)+\sum_{\alpha\in\Gamma^{(-)}}(\hat{a}^{(\alpha)\dagger}_{\omega}\Phi^{(\alpha)}_\omega+\hat{a}^{(\alpha)}_{\omega}\sigma_1\Phi^{(\alpha)*}_\omega)\Bigg],
\end{align} 
and if we write $\Psi^{(\alpha)}_{\omega}= \exp(-i\omega t)\left(\begin{matrix}f^{(\alpha)}_{\omega}\\ h^{(\alpha)}_{\omega}\end{matrix}\right)$, 
we finally obtain
\begin{align}
\psi=&\sqrt{\rho}\int_0^\infty\d\omega\Bigg[ \sum_{\alpha\in\Gamma^{(+)}}\left(\hat{a}^{(\alpha)}_{\omega}e^{-i\omega t/\xi^2_{\rm u}}f^{(\alpha)}_\omega+\hat{a}^{(\alpha)\dagger}_{\omega}e^{i\omega t/\xi^2_{\rm u}}h^{(\alpha)*}_\omega\right)+\sum_{\alpha\in\Gamma^{(-)}}\left(\hat{a}^{(\alpha)\dagger}_{\omega}e^{-i\omega t/\xi^2_{\rm u}}f^{(\alpha)}_\omega+\hat{a}^{(\alpha)}_{\omega}e^{i\omega t/\xi^2_{\rm u}}h^{(\alpha)*}_\omega\right)\Bigg].
\end{align} 

\subsection{Local versus global energy conservation}

From the Lagrangian \eqref{lagrangian} we calculate the canonically conjugate momentum $\pi=\delta L_{\psi}/\delta (\partial_t\psi)=i\psi^*/2$, and the Hamiltonian $H=\int dx(\pi\partial_t\psi+\pi^*\partial_t\psi^*)-L_{\psi}=\int dx \mathcal{H}$, where
\begin{align}
\mathcal{H}=\frac{1}{2}|\partial_x\psi|^2+\left(g_{\rm dd}\rho+\frac{\partial_x^2\sqrt{\rho}}{2\sqrt{\rho}}\right)|\psi|^2+\frac{g_{\rm dd}}{2}\rho(\psi^{*2}+\psi^{2})-\frac{iv}{2}\left[\psi^*\partial_x\psi-(\partial_x\psi^*)\psi\right]-\frac{3g_{\rm dd}}{2}\sqrt{\rho}(\psi+\psi^*)G*\sqrt{\rho}(\psi+\psi^*).\label{hamiltoniandensity}
\end{align}
From the Hamiltonian density \eqref{hamiltoniandensity} we obtain
\begin{equation}
\partial_t\mathcal{H}=-\partial_x S+\frac{3g_{\rm dd}}{2}\sqrt{\rho}[\partial_t(\psi+\psi^*)]G*\sqrt{\rho}(\psi+\psi^*)-\frac{3g_{\rm dd}}{2}\sqrt{\rho}(\psi+\psi^*)G*\sqrt{\rho}\partial_t(\psi+\psi^*),
\end{equation}
where $S=(-1/2)\{(\partial_t\psi^*)(\partial_x+iv)\psi+[(\partial_x-iv)\psi^*]\partial_t\psi\}$. We thus observe that unless $g_{\rm dd}=0$ or $\beta=0$ the system energy is not locally conserved in general. Nevertheless, the system total energy in its ground state is still conserved. By performing this integral and making the substitution $\omega'\rightarrow\omega$, $\alpha'=\alpha$, we obtain the constraint
%
%

{Indeed,} the system energy in its ground state is given by $H=\int \d x\langle\hat{\mathcal{H}}\rangle$, and the Hamiltonian operator $\hat{\mathcal{H}}$ is obtained from Eq.~\eqref{hamiltoniandensity} by making $\psi\rightarrow \hat{\psi}$ followed by normal ordering. Now, because of stationarity we have $\partial_t H=0$, which follows from $\partial_t \langle\hat{\mathcal{H}}\rangle=0$.  The condition $\partial_t H=0$ then has a clear physical meaning: If the system radiates, the power emitted at $x\rightarrow-\infty$ equals the power absorbed at $x\rightarrow\infty$. Our goal is to calculate the radiation power $S_{-\infty}$ at  $x\rightarrow-\infty$, and thus $\partial_t H=-(S_{\infty}-S_{-\infty})=0$. We find that
\begin{align}
\partial_t \langle\hat{\mathcal{H}}\rangle= 
i\int_{0}^{\infty}\d\omega \omega\bigg\{&\bigg(\sum_{\alpha\in\Gamma^{(+)}}-\sum_{\alpha\in\Gamma^{(-)}}\bigg)\bigg[\frac{\partial_x}{2}\rho\bigg(h^{(\alpha)*}_{\omega}\partial_xh^{(\alpha)}_{\omega}-h^{(\alpha)}_{\omega}\partial_xh^{(\alpha)*}_{\omega}-2i\mathfrak{m}_{\rm u}\frac{\rho_{\rm u}}{\rho}|h^{(\alpha)}_{\omega}|^2\bigg)\nonumber\\
&+\frac{3\rho}{\rho_{\rm u}}h^{(\alpha)*}_{\omega}G*\rho(f^{(\alpha)}_{\omega}+h^{(\alpha)}_{\omega})+\frac{3\rho}{\rho_{\rm u}}h^{(\alpha)}_{\omega}G*\rho(f^{(\alpha)*}_{\omega}+h^{(\alpha)*}_{\omega})\bigg]\bigg\}
+ 
i\int_{0}^{\infty}\d\omega \omega
\sum_{\alpha\in\Gamma^{(-)}}I^{\alpha,\alpha}_{\omega,\omega}.\label{eqenergy}
\end{align} 
From Eq.~\eqref{eqenergy}, because $\partial_t \langle\hat{\mathcal{H}}\rangle=0$ and $I^{\alpha,\alpha}_{\omega,\omega}=0$, we conclude that the first term inside the curly brackets when integrated over $\omega$ gives 0 for all $x$, and thus the net contribution of this term to $S_{-\infty}$ is zero.  Accordingly, the second term, which is zero 
irrespective of the integration  in $\omega$, gives rise to the radiation power. 
Indeed, $I^{\alpha,\alpha}_{\omega,\omega}=0$ contains the gradient of a constant function containing the radiated power, and which can be determined as follows. Note that 
\begin{equation}
\partial_t H=\partial_t\int_{-\infty}^{\infty}\d x\langle\mathcal{H}\rangle=
i\int_{0}^{\infty}\d\omega \omega\sum_{\alpha\in\Gamma^{(-)}}\int_{-\infty}^{\infty}\d x I^{\alpha,\alpha}_{\omega,\omega},
\end{equation}
and $\int_{-\infty}^{\infty}\d x I^{\alpha,\alpha}_{\omega,\omega}$ can be calculated with the aid of Eq.~\eqref{cons1} taking $x'\rightarrow\infty$ in
\begin{align}
i\int_{-x'}^{x'}\d xI^{\alpha,\alpha}_{\omega,\omega'}=&\rho_{\rm u}\sum_{k,k'}S^{(\alpha)*}_{k}S^{(\alpha)}_{k'}\Phi^\dagger_{k}\left[\frac{k'+k^*}{2}+\mathfrak{m}_{\rm u}\sigma_3-3\sigma_4\frac{\tilde{G}(\beta k')-\tilde{G}(\beta k^*)}{k'-k^*}\right]\Phi_{k'}e^{-ix'(k'-k^*)}\nonumber\\
&-\rho_{\rm d}\sum_{p,p'}S^{(\alpha)*}_{p}S^{(\alpha)}_{p'}\Phi^\dagger_{p}\left[\frac{p'+p^*}{2}+\mathfrak{m}_{\rm u}\frac{\rho_{\rm u}}{\rho_{\rm d}}\sigma_3-3\frac{\rho_{\rm d}}{\rho_{\rm u}}\sigma_4\frac{\tilde{G}(\beta p')-\tilde{G}(\beta p^*)}{p'-p^*}\right]\Phi_{p'}e^{ix'(p'-p^*)},
\end{align}
where we used the orthogonality condition $\int_{-\infty}^{\infty}\d x I^{\alpha,\alpha}_{\omega,\omega'}=0$ and the sums in primed wave vectors refer to $\omega'$, whereas unprimed wave vectors correspond to $\omega$. Thus, by taking $\omega'\rightarrow \omega$ and $x'\rightarrow \infty$ we obtain
\begin{align}
\partial_tH= 
-\frac{1}{2\pi}
\int_{0}^{\infty}\d\omega \omega\sum_{\alpha\in\Gamma^{(-)}}\bigg\{\sum_{p\ {\rm prop}}|S^{(\alpha)}_{p}|^2\sgn\left[V(p)(\omega-\mathfrak{m}_{\rm u}p\rho_{\rm u}/\rho_{\rm d})\right]-\sum_{k\ {\rm prop}}|S^{(\alpha)}_{k}|^2\sgn\left[V(k)(\omega-\mathfrak{m}_{\rm u}k)\right]\bigg\},
\end{align} 
and {the outgoing flux at upstream infinity becomes} 
\begin{align}
S_{-\infty}=
\frac{1}{2\pi}
\int_{0}^{\infty}\d\omega \omega\sum_{\alpha\in\Gamma^{(-)}}\sum_{k\ {\rm prop}}|S^{(\alpha)}_{k}|^2\sgn\left[V(k)(\omega-\mathfrak{m}_{\rm u}k)\right].
\end{align} 
{Note, in particular, that the unitarity condition \eqref{smatrix} for $\alpha'=\alpha$ implies that $S_{\infty}$=$S_{-\infty}$, i.e., global energy conservation.}

\end{widetext}
\end{document}